\documentclass[11pt,a4paper]{article}

\usepackage{subfigure,jcappub,amsfonts,amsmath,amssymb,bm,graphicx}

\usepackage[active]{srcltx}

\title{Birefringence of electromagnetic waves in the relic neutrino gas}

\author{Maxim Dvornikov and}
\emailAdd{maxdvo@izmiran.ru}
\author{Victor B. Semikoz}
\affiliation{Pushkov Institute of Terrestrial Magnetism, Ionosphere
and Radiowave Propagation (IZMIRAN),
108840 Moscow, Troitsk, Russia}
\emailAdd{semikoz@yandex.ru}

\abstract{
We reconsider the problem of the birefringence of electromagnetic (EM) waves  in a medium consisting of a plasma and a $\nu\bar{\nu}$-gas within the Standard Model of particle physics. The considered effect arises in such a medium due to the parity violation for the electroweak neutrino-electron interaction. Our recent calculations of the electroweak correction to the photon polarization operator in the electroweak plasma allow us to significantly improve some previous estimates of such effect in astrophysics. We estimate the rotary power for EM waves propagating in a non-relativistic plasma in the intergalactic space and interacting with the gas of relic neutrinos and antineutrinos there. We show that, in presence of a plasma, the EM wave birefringence effect in a $\nu\bar{\nu}$-gas exceeds significantly that effect in a $\nu\bar{\nu}$-gas in empty space considered earlier. These previous treatments of the birefringence relied on the calculations of the refraction index for on-shell photons in vacuum using the forward scattering amplitude $\gamma\nu\to \gamma\nu$ with virtual charged leptons in Feynman diagrams. The possibility to observe experimentally the new effect suggested here is discussed.
}
\begin{document}

\maketitle

%\section{Introduction}
\section{Introduction\label{INTR}}
The detection of the Cosmic Neutrino Background (C$\nu$B) is still an actual problem in particle physics and astrophysics.
Unfortunately, the direct detection through either neutrino capture on beta decaying nuclei \cite{Weinberg,Cocco:2009rh}, in particular,
through a C$\nu$B capture by tritium in the PTOLEMY experiment \cite{Betti:2019ouf}, or using other methods, e.g.,
measuring a torque of a heavy ferromagnet moving through neutrino-antineutrino sea \cite{Stodolsky:1974aq} remains an intriguing hard to get goal done. On the other hand, indirect methods based on cosmological and astrophysics data, e.g., using CMB anisotropy measurements in WMAP and Plank experiments \cite{Hannestad}, could be more promising to detect C$\nu$B. 
Let us stress the importance of a degeneracy of the relic neutrino (antineutrino) gas since any macroscopic effect, involving C$\nu$B, should be  proportional to the density difference $(n_{\nu} - n_{\bar{\nu}})$. 

A promising ability comes from the propagation of electromagnetic waves through the neutrino-antineutrino sea if we take into account the Chern-Simons (CS) anomaly term $\mathcal{L}_\mathrm{CS}=\Pi_2({\bf A}\cdot{\bf B})$, where ${\bf A}$ is the vector potential and ${\bf B}$ is the magnetic field, entering the effective Lagrangian density in the standard model (SM) for the photon polarization operator in electroweak plasma. The detailes of the derivation of $\mathcal{L}_\mathrm{CS}$ can be found in refs.~\cite{Boyarsky:2012ex,Dvornikov:2013bca}. The form-factor $\Pi_2$, which will be discussed in section~\ref{sec:POLAROPER}, defines the induced pseudovector current ${\bf j}_5=\Pi_2{\bf B}$ in the interaction Lagrangian term $\mathcal{L}_\mathrm{CS}={\bf j}_5{\bf A}$ that violates parity.
Below we concentrate on the corresponding birefringence effect for electromagnetic waves in such an appropriate medium.

There were numerous attempts to calculate the photon polarization operator $\Pi_{\mu\nu}(\omega, \mathbf{k})$, where $\omega$ is the photon frequency and  $\mathbf{k}$ is its wave vector, in the neutrino sea 
using the neutrino-photon forward scattering amplitude in vacuum, or the refraction index for on-shell photons propagating in  $\nu\bar{\nu }$ gas, see e.g. ref.~\cite{Langacker:1992xk}. For on-shell photons and massless neutrinos the point-like weak interaction leads to the vanishing amplitude for photon-neutrino interactions according to the Gell-Mann theorem~\cite{Gell-Mann}, when one uses the triangle diagram for $\nu e$-scattering. Thus, the antisymmetric term in the polarization tensor vanishes, $\Pi^{(\nu e)}_{\mu\nu}\to 0$, in the lowest order $\sim \alpha_\mathrm{em}G_\mathrm{F}$, where $\alpha_\mathrm{em}=e^2/4\pi=1/137$ is the fine structure constant and $G_\mathrm{F}=1.17\times 10^{-5}\,\mathrm{GeV}^{-2}$ is the Fermi constant. To escape the Gell-Mann's restriction, one needs to consider non-local interactions in order to include higher orbital momenta of a $\nu\bar{\nu}$ pair into the annihilation process  instead of $L=0$ for such a pair in the point-like approximation~\cite{Gell-Mann}. It can be achieved by using the exact $W$-boson propagator that changes a loop from a triangle diagram to a square one, with the gauge boson providing one of the sides~\cite{Novikov}.

The corresponding effect of the birefringence of electromagnetic waves propagating in vacuum filled by a neutrino-antineutrino gas, estimated in ref.~\cite{Novikov}, is not zero. The rotary power, considered in section~\ref{sec:ROTARY} below, is of the order $\phi/l\sim \alpha_\mathrm{em}(n_{\nu} - n_{\bar{\nu}})/M_\mathrm{W}^6$, where $M_\mathrm{W}$ is the $W$-boson mass and $\phi$ is the rotation angle of the photon polarization vector which whirls in the $(x,y)$ plane perpendicular to the photon momentum at the distance $z=l$ from a source \cite{Novikov,Abbasabadi:2001ps,Abbasabadi:2003uc}. That result
corrects the estimate $\phi/l\sim \alpha_\mathrm{em}G_\mathrm{F}k_\mathrm{F}^3$ obtained in ref.~\cite{Royer}, where on shell photons, scattered off neutrinos in a degenerate neutrino sea, were discussed. The additional suppression factor $(k_\mathrm{F}\omega)^2/M_\mathrm{W}^4$ was found in ref.~\cite{Novikov} to be present in the rotary power with the outcome $\phi/l\sim \alpha_\mathrm{em}G_\mathrm{F}k_\mathrm{F}^3(k_\mathrm{F}\omega)^2/M_\mathrm{W}^4$.

Nevertheless, for the off-shell photons in the case of the neutrino-antineutrino sea embedded into plasma the rotary power can be linear over Fermi constant, $\phi/l\sim \alpha_\mathrm{em}G_\mathrm{F}(n_{\nu} - n_{\bar{\nu}})$. Note that such a result was not completed in ref.~\cite{Mohanty:1997mr} since the factor $q^2=0$ was formally changed to the transverse plasmon mass $q^2\equiv \omega^2 - {\bf k}^2=\omega^2_p$, where $\omega_p^2=e^2n_e/m_e$ is the plasma frequency. In ref.~\cite{Mohanty:1997mr}, a neutrino-photon forward scattering amplitude with free charged fermion propagators was used in the calculation of the parity violating part of the polarization tensor $\Pi_{\mu\nu}$. This approximation is appropriate for a neutrino sea filling vacuum while it is not sufficient in the case of neutrinos in a plasma. For that case, using the results of ref.~\cite{Dvornikov:2013bca}, we rely below on the exact calculation of the parity odd term in the polarization operator accounting for electroweak $\nu e$ interactions in plasma with real charged leptons (electrons).

Our work is organized as follows. In section~\ref{sec:POLAROPER}, we discuss parity-odd terms in the photon polarization operator $\Pi_{\mu\nu}$ for the neutrino sea embedded
into vacuum and in a plasma. In section~\ref{sec:MAXWELL}, we modify Maxwell equations in SM accounting for the corresponding CS term in the effective Lagrangian for electromagnetic fields, $\mathcal{L}_\mathrm{CS}$. In section~\ref{sec:BF}, we derive the dispersion relation for photons in the presence of such CS term  and obtain their birefringence resulting in the corresponding rotary power $\Phi=\phi/l$ at the distance $l$ from a remote source. Then, in section~\ref{sec:ROTARY}, we compare different scenarios for the birefringence effect illustrated in figure~\ref{fig:rotarypower}: a) calculations using Feynmann diagrams for $\nu\gamma$ forward scattering amplitudes in vacuum; b) the Faraday effect in InterGalactic Magnetic Fields (IGMF's), and c) our approach using direct calculations of the parity-odd term in the tensor $\Pi_{\mu\nu}$ in a plasma that is based on the Dirac equation for electrons accounting for their interactions with the $\nu\bar{\nu}$ gas in SM. Finally, in section~\ref{sec:DISC}, we discuss outcomes for the birefringence effect. We review briefly our approach based on the Dirac equation and the polarization tensor $\Pi_{\mu\nu}$ in appendix~\ref{Dirac}. 

\section{Polarization operator and Chern-Simons term in the effective Lagrangian\label{sec:POLAROPER}} 

In ref.~\cite{Dvornikov:2013bca}, we calculated the relevant parity violating part of the polarization operator, $\Pi_{\mu\nu}(q)=\mathrm{i}\varepsilon_{\mu\nu\alpha\beta}q^{\alpha}(f_\mathrm{L}^{\beta}-f_\mathrm{R}^{\beta})\Pi_\mathrm{P}(q^2)$, which depends on the sum of vacuum ($T=0,\mu=0$) and the plasma temperature terms ($T\neq 0$ , $\mu\neq 0$), $\Pi_\mathrm{P}(q^2)=\Pi_\mathrm{P}^{(\mathrm{vac})} + \Pi_\mathrm{P}^{(\mathrm{T})}$. Here $\mu$ is the chemical potential of charged leptons, $f^{\beta}_\mathrm{L} - f^{\beta}_\mathrm{R}=(f_\mathrm{L}^0 -f_\mathrm{R}^0)u^{\beta}$ is given by the unit fluid four-velocity $u^{\beta}$, $u^{\beta}u_{\beta}=1$. The 
axial part of the weak $\nu e$ interaction potential in the Dirac equation,  $V_\mathrm{A}=-(f_\mathrm{L}^0 - f_\mathrm{R}^0)/2$, is present in  eq.~(\ref{potentials}) in appendix~\ref{Dirac}. The factor
\begin{equation}\label{difference}
f_\mathrm{L}^0 - f_\mathrm{R}^0=2\sqrt{2}G_\mathrm{F}\left[\Delta n_{\nu_e} - \frac{1}{2}\sum_{\alpha}\Delta n_{\nu_{\alpha}}\right],
\quad
\alpha=e,\mu,\tau,
\end{equation}
includes both the charged current (CC) and neutral current (NC) contributions in the electroweak $\nu e$ interactions where we use the point-like Fermi approximation, and $\Delta n_{\nu_{\alpha}}= n_{\nu_{\alpha}} - n_{\bar{\nu}_{\alpha}}$ are the neutrino number density asymmetries.

If we confine our problem with the CC contribution to the weak $\nu e$ interaction potential, $V_\mathrm{A}=-\sqrt{2}G_\mathrm{F}(n_{\nu_e} - n_{\bar{\nu}_e})$, then for a soft photon, $q^2< 4m_e^2$, the vacuum part calculated in ref.~\cite{Dvornikov:2013bca} with virtual charged lepton propagators equals to 
\begin{equation}\label{Mohanty}
\Pi_\mathrm{P}^{(\mathrm{vac})}(q^2)= - \frac{\alpha_\mathrm{em}}{\pi}\left(\frac{q^2}{m_e^2}\right)\int_0^1\mathrm{d}x\frac{x(1-x)}{1 - x(1 -x)(q^2/m_e^2)}.
\end{equation}
Accounting for the factor in eq.~(\ref{difference}) in the 3-tensor part of $\Pi_{\mu\nu}$, $\Pi_{ij}=\mathrm{i}\varepsilon_{ijn}k^n(f^0_\mathrm{L} - f_\mathrm{R}^0)\Pi_\mathrm{P}(q^2)$, eq.~\eqref{Mohanty} leads
 in the limit, $q^2\ll m_e^2$,\footnote{For the plasma frequency $\omega_p=5.65\times 10^4\sqrt{n_e}\,\text{s}^{-1}=6.5\times 10^{-13}\,\mathrm{eV}$, given by a small $n_e\sim 3\times 10^{-4}\,\text{cm}^{-3}$ in the galo of a Milky-Way-like galaxy \cite{Hammond:2012pn} and for the transversal plasmon $q^2=\omega_p^2$, such a limit, $q^2\ll m_e^2$, is fulfilled.} to the exact coincidence with eq.~(3.15) in ref.~\cite{Mohanty:1997mr} for $\Pi_2^{(\mathrm{vac})}(q^2)=(f_\mathrm{L}^0 - f_\mathrm{R}^0)\Pi_\mathrm{P}^{(\mathrm{vac})}(q^2)$:
\begin{equation}\label{Mohanty2}
\Pi_2^{(\mathrm{vac})}(q^2)=\frac{\sqrt{2}G_\mathrm{F}\alpha_\mathrm{em}}{3\pi}\left(\frac{q^2}{m_e^2}\right)(n_{\nu_e} - n_{\bar{\nu}_e}).
\end{equation}
Obviously, this term vanishes for photons on-shell, $q^2=0$, obeying Gell-Mann theorem \cite{Gell-Mann}. Thus, there is no a reason to consider a birefringence for a neutrino sea embedded into empty space~\cite{Mohanty:1997mr}.

In the presence of a nonrelativistic plasma and $\nu\bar{\nu}$ -gas in the intergalactic space when accounting for both CC and NC contributions to $\Pi_2^{(\nu )}=(f_\mathrm{L}^0 - f_\mathrm{R}^0)\Pi_\mathrm{P}^{(\mathrm{vac})}(q^2)$, one gets instead of eq.~(\ref{Mohanty2}),
\begin{equation}\label{Mohanty3}
\Pi_2^{(\nu )}=-\frac{2\alpha_\mathrm{em}^2}{3}(f_\mathrm{L}^0 - f_\mathrm{R}^0)\frac{n_e}{m_e^3},
\end{equation}
where we substitute the transversal plasmon spectrum, $q^2=\omega_p^2=4\pi\alpha_\mathrm{em}n_e/m_e\neq 0$ in the vacuum term in eq.~(\ref{Mohanty}). Let us stress that charged leptons in propagators are virtual in eq.~(\ref{Mohanty3}). Neither a temperature nor the chemical potential in plasma are taken into account.

The temperature term $\Pi_2^{(\mathrm{T})}=(f_\mathrm{L}^0 - f_\mathrm{R}^0)\Pi_\mathrm{P}^{(\mathrm{T})}$ in the total sum $\Pi_2=\Pi_2^{(\nu )}(q^2) + \Pi_2^{(\mathrm{T})}$ calculated in ref.~\cite{Dvornikov:2013bca} using the Matsubara technique~\cite{Kapusta} for the case of the classical electron gas distribution,\footnote{The plasma density is given by the standard integral $n_e=g_e\smallint f(\varepsilon_{\bf{p}}) \mathrm{d}^3p/(2\pi)^3,$ where $g_e=2$ for electrons.} $f(\varepsilon_{\bf{p}})=\exp[\beta(\varepsilon_{\bf{p}} - \mu)]\gg 1$ and in the limit  ${\rm max}\{\omega^2,{\bf k}^2\}\ll m_e^2$,  reduces to
\begin{equation}\label{Matsubara}
\Pi_2^{(\mathrm{T})}= - \frac{7}{6}e^2(f_\mathrm{L}^0 - f_\mathrm{R}^0)\int\frac{\mathrm{d}^3p}{(2\pi)^3\varepsilon_{\bf{p}}^3}\left(\frac{m_e^2}{\varepsilon_{\bf{p}}^2} + \frac{m_e^2\beta}{\varepsilon_{\bf{p}}} - \frac{\beta^2{\bf p}^2}{3}\right)\exp[\beta(\mu - \varepsilon_{\bf{p}})],
\end{equation}
where $\varepsilon_{\bf p}=\sqrt{{\bf p}^2 + m_e^2}$ is the electron energy, $\beta=T_e^{-1}$ is the reciprocal temperature, $\mu=m_e + T_e\ln [n_e(2\pi)^{3/2}/g_e(m_eT_e)^{3/2}]$ is the chemical potential of the classical electron gas.

For a nonrelativistic electron gas, $T_e\ll m_e$, the main contribution in eq.~(\ref{Matsubara}) comes from the first term in integrand  ($\sim m_e^2/\varepsilon_{\bf p}^2$). It means that the plasma term $\Pi_2^{(\mathrm{T})}$ takes the form,\footnote{Note that the last terms in integrand in eq.~(\ref{Matsubara}), $m_e^2\beta/\varepsilon_{\bf p} - \beta^2{\bf p}^2/3$, being integrated, cancel each other, i.e. their sum equals to zero, for the non-relativistic Boltzmann distribution.}
\begin{equation}\label{nonrel}
\Pi_2^{(\mathrm{T})}=- \frac{7\pi\alpha_\mathrm{em}}{3}(f_\mathrm{L}^0 - f_\mathrm{R}^0)\frac{n_e}{m_e^3}.
\end{equation}
The result in eq.~\eqref{nonrel} is irrelevant to the claim of the Gell-Mann theorem \cite{Gell-Mann} for which the parity violation term $\Pi_2$ should vanish  in the lowest order in Fermi constant $\sim G_\mathrm{F}$, $\Pi_2=0$. In the case for $\Pi_2^{(\mathrm{T})}$ in eq.~(\ref{nonrel}), charged leptons are real particles in plasma rather than virtual ones in propagators as in the case of eqs.~(\ref{Mohanty2}) and~(\ref{Mohanty3}). For a dense relativistic plasmas, the particle energy $\varepsilon_{\bf p}=\sqrt{{\bf p}^2 + M^2}$ depends on the effective mass $M^2=m^2 -q^2x(1-x)$ in propagators~\cite{Dvornikov:2013bca}. Moreover, the photon's dispersion relation differs from the vacuum one, $q^2=k_0^2 - {\bf k}^2\neq 0$~\cite{Braaten:1993jw}. These facts are contrary to calculations of one-loop Feynman diagrams for the neutrino-photon forward scattering in ref.~\cite{Mohanty:1997mr}, where photons should be on-shell, $q^2=0$. For a low density plasma, present in intergalactic space, with photon eigenmodes inside it, one has that ${\rm max}(k_0^2,{\bf k}^2)\ll m^2$ and $q^2\neq 0$. The particles become free, $\varepsilon_{\bf p}=\sqrt{{\bf p}^2 + m^2}$. It means that, in a non-relativistic classical electron gas, e.g., in intergalactic space, where $n_e\simeq 10^{-3}\,\text{cm}^{-3}$ and $T\sim 5\,\mathrm{keV}\ll m_e $, the parity violation term  in eq.~(\ref{nonrel}) differs from zero, $\Pi_2^{(\mathrm{T})}\neq 0$, being beyond the scope of the Gell-Mann theorem. 

Thus, the three vector part of the induced electric current $j^{(\text{ind})}_{\mu}=\Pi_{\mu\nu}A^{\nu}$ in SM plasma contains the parity violation term ${\bf j}_5(\omega,{\bf k})=\Pi_2{\bf B}(\omega,{\bf k})$ coming from the pseudovector current in the $\nu e$ interaction, where, in the total $\Pi_2=\Pi^{(\nu )}_2 + \Pi_2^{(\mathrm{T})}\approx \Pi_2^{(\mathrm{T})}$, we neglect the term with virtual charged leptons in eq.~(\ref{Mohanty3}) since $\Pi_2^{(\mathrm{T})}/\Pi_2^{(\nu )}=7\pi /(2\alpha_\mathrm{em})\sim 10^3\gg 1$. Notice that, substituting $q^2=\omega^2 - {\bf k}^2=\omega_p^2$ for transverse plasmons into eq.~(\ref{Mohanty2}), we consider the $\nu\bar{\nu}$-gas embedded into plasma. Such substitution is allowed for both terms, $\Pi_2^{(\mathrm{vac})}$ and $\Pi_2^{(\mathrm{T})}$, contrary to the case in ref.~\cite{Mohanty:1997mr}. The corresponding Chern-Simons term in the effective Lagrangian density for electromagnetic waves takes the form $\mathcal{L}_\mathrm{CS}=\Pi_2({\bf A}\cdot{\bf B})\approx \Pi_2^{(\mathrm{T})}({\bf A}\cdot{\bf B})$ that is proportional to the magnetic helicity density $h=({\bf A}\cdot{\bf B})$.

\section{Maxwell equations generalized in SM plasma\label{sec:MAXWELL}}

In an isotropic plasma accounting for electroweak interactions in SM, Maxwell equations in the Fourier  representation take the form~\cite{Nieves:1992et},
\begin{eqnarray}\label{Maxwell} 
&&{\bf k}\cdot{\bf B}=0,
\quad
{\bf k}\times{\bf E}=\omega {\bf B},
\quad
\mathrm{i}\epsilon{\bf k}\cdot{\bf E}=\rho_\mathrm{ext},\nonumber\\&&
\mu^{-1}\mathrm{i}{\bf k}\times {\bf B} + \mathrm{i}\omega \epsilon {\bf E} + \mathrm{i}\zeta\omega {\bf B}={\bf j}_\mathrm{ext},
\end{eqnarray}
where $\epsilon$ is the dielectric permittivity, $\mu$ is the magnetic permeability, and $\zeta$ is the chiral dispersion characteristic in plasma. In SM, for an isotropic medium, this third dimensionless constant, describing the electromagnetic properties of plasma (in addition to the two standard ones: $\epsilon$ and $\mu$), should obey some C, P, and T symmetry properties. The requirement that the fields ${\bf E}$ and ${\bf B}$ and the current density ${\bf j}$ are real quantities in the coordinate space, as well as the time reversal symmetry $\zeta (k, - \omega)= - \zeta (k, \omega)$, implies that $\zeta$ is purely imaginary and odd function of $\omega$~\cite{Nieves:1992et}.

Note that, in isotropic plasma, one can put $\mu=1$ since, in the high frequency limit $\omega\gg k \langle v \rangle$, one can neglect a spatial dispersion. As a result, the general permittivity tensor $\epsilon_{ij}(\omega,k)=(\delta_{ij} - k_ik_j/k^2)\epsilon^{(\mathrm{tr})}(\omega,k) + \epsilon^{(\mathrm{l})}(\omega,k)k_ik_j/k^2$ for $\epsilon^{(\mathrm{tr})}(\omega)=\epsilon^{(\mathrm{l})}(\omega)=1 - \omega_p^2/\omega^2=\epsilon (\omega)$ takes the
simple form, $\epsilon_{ij}=\delta_{ij}\epsilon (\omega)$, where $\omega_p^2=e^2n_e/m_e$ is the plasma frequency. Then, the well-known relation 
\begin{equation}\label{permeability}
1 - \frac{1}{\mu (\omega, k)}=\frac{\omega^2}{k^2}\left[\epsilon^{(\mathrm{tr})}(\omega, k) - \epsilon^{(\mathrm{l})}(\omega,k)\right]\to 0
\end{equation}
gives $\mu=1$. Thus, we use below the two dispersion characteristics $\epsilon (\omega)$ and $\zeta (\omega)$ obeying the known symmetry properties. Our main goal is to find the new chiral parameter $\zeta (\omega)$ in SM. Note that parity conservation in QED plasma automatically gives $\zeta=0$, while extending a model to SM one can expect $\zeta\neq 0$. 

In the absence of external currents and charges, $\rho_\mathrm{ext}={\bf j}_\mathrm{ext}=0$, accounting for the additional pseudovector current ${\bf j}_5=\Pi_2{\bf B}(\omega,{\bf k})$ given by the CS term $L_\mathrm{CS}=\Pi_2^{(\mathrm{T})}({\bf A}{\bf B})$ in effective Lagrangian, we get the modified Maxwell equation generalized in SM due to the parity violation, 
\begin{equation}\label{Maxwell3}
\mathrm{i}\omega {\bf E}(\omega,{\bf k}) + \mathrm{i}{\bf k}\times {\bf B}(\omega,{\bf k})={\bf j}_\mathrm{ind}(\omega,{\bf k})+ \Pi_2^{(\mathrm{T})}{\bf B}(\omega,{\bf k}) .
\end{equation} 
In the right hand side of eq.~(\ref{Maxwell3}), the induced vector current ${\bf j}_\mathrm{ind}(\omega,{\bf k})=\sigma_\mathrm{cond}(\omega){\bf E}(\omega,{\bf k})=-\mathrm{i}[\epsilon(\omega) - 1]\omega{\bf E}(\omega,{\bf k})$ is the standard ohmic current in the correspondence with the relation of the dielectric permittivity and the conductivity in an isotropic plasma, $\epsilon (\omega)=1 +\mathrm{i}\sigma_\mathrm{cond}(\omega)/\omega$. Substituting this vector current into eq.~(\ref{Maxwell3}) one can recast the second line in eq.~(\ref{Maxwell}) as
\begin{equation}\label{Maxwell4}
\mathrm{i}{\bf k}\times {\bf B}(\omega,{\bf k}) + \mathrm{i}\omega \epsilon (\omega){\bf E}(\omega,{\bf k}) + \mathrm{i}\zeta(\omega)\omega {\bf B}(\omega,{\bf k})= 0,
\end{equation}
where the third dispersion characteristic $\zeta$, \begin{equation}\label{zeta}
\zeta(\omega)=\frac{\mathrm{i}\Pi_2^{(\mathrm{T})}}{\omega},
\end{equation} takes the explicit form, being purely imaginary and the odd function of $\omega$, as it should be.

\section{Birefringence of electromagnetic waves in the isotropic SM plasma\label{sec:BF}}

For the intergalactic plasma density in galactic clusters $n_e\simeq 10^{-3}\,\text{cm}^{-3}$ filled by hot electrons with $T_e\sim 5\,\text{keV}$, one obtains $\omega_p=1.2\times 10^{-12}\,\text{eV}$. It leads to a small rotary power (see below). 
Substituting ${\bf B}=({\bf k}\times {\bf E})/\omega$ one can easily get from eq.~(\ref{Maxwell4}) the dispersion equation for the right and left circularly polarized states of transverse electromagnetic waves, ${\bf k}\cdot{\bf E}=0$ and ${\bf E}=E(\omega,k)\hat{e}_{\pm}$, 
\begin{equation}\label{dispersion}
\omega^2 -k^2\left(\frac{1}{\epsilon} \pm \frac{\mathrm{i}\omega\zeta}{\epsilon k}\right)=0,
\end{equation}
where
\begin{equation}
\hat{e}_{\pm}=\frac{1}{\sqrt{2}}({\bf e}_1 \pm \mathrm{i}{\bf e}_2), ~~{\bf e}_2={\bf k}\times{\bf e}_1/k, 
\end{equation}
are the right and left polarization vectors, $E(\omega,k)$ is the wave amplitude. Substituting the electroweak chiral parameter in eq.~(\ref{zeta}), $\Pi_2^{(\mathrm{T})}\ll (\omega, k)$, one obtains from the dispersion eq.~(\ref{dispersion}) the two dispersion relations $\omega_{\pm}(k)$ for a fixed wave number $k$,
\begin{equation}\label{frequency}
\omega_{\pm}=\sqrt{\omega_p^2 + k^2} \mp \left(\frac{\Pi_2^{(\mathrm{T})}}{2}\right)\frac{k}{\sqrt{\omega^2_p + k^2}},
\end{equation}
or the two wave numbers,
\begin{equation}\label{wavenumber}
k_{\pm}= \sqrt{\omega^2 - \omega_p^2} \pm \frac{\Pi_2^{(\mathrm{T})}}{2}.
\end{equation}
for a fixed frequency $\omega$.

One of the issues of the optical activity in media is the rotation of polarization vector in the plane perpendicular to the direction of wave propagation. Choosing the $z$-axis parallel to the photon momentum, ${\bf k}=(0,0,k)$, we can treat at the source position (point $z=0$) a plane polarized wave as an equal admixture of right and left circularly polarized waves,
\begin{align}\label{admixture}
  {\bf E}(z,t) =& E_{\omega}e^{-\mathrm{i}\omega t}\frac{1}{\sqrt{2}}\left(e^{\mathrm{i}k_+z}\hat{e}_+
  + e^{\mathrm{i}k_-z}\hat{e}_-\right)
  \notag
  \\
  & = E_{\omega}e^{-\mathrm{i}\omega t}
  e^{\mathrm{i}(k_+ +k_-)z/2}
  \frac{1}{\sqrt{2}}
  \left(e^{\mathrm{i}(k_+ -k_-)z/2}\hat{e}_+ + e^{-\mathrm{i}(k_+ -k_-)z/2}\hat{e}_-\right).
\end{align}
Here, at the point $z=0$ (source position), the polarization vector is directed along ${\bf e}_1$, chosen as the $x$-axis. Then it rotates in the $(x,y)$-plane and points at the distance $z=l$ at the angle (relative to the $x$-axis) given by 
\begin{equation}\label{angle}
\phi (l)=\frac{1}{2}\left(k_+ - k_-\right)l=\Pi_2^{(\mathrm{T})}l,
\end{equation}
where factor $\Pi_2^{(\mathrm{T})}$ is given by eq.~(\ref{nonrel}).
Let us discuss some applications and compare predictions of different models of the birefringence.

\section{Comparison of rotary power with previous predictions\label{sec:ROTARY}}

It is instructive to compare the results of different calculations for the rotary power $\Phi=\phi/l$. For a photon with the four-momentum $q^{\mu}=(\omega,{\bf k})$, propagating in vacuum ($q_{\mu}q^{\mu}=0$) filled by the neutrino-antineutrino sea, one obtains \cite{Abbasabadi:2001ps,Abbasabadi:2003uc}:
\begin{equation}\label{Repko}
\frac{\phi}{l}=\frac{112\pi G_\mathrm{F}\alpha_\mathrm{em}}{45\sqrt{2}}\left[\ln \left(\frac{M_\mathrm{W}}{m_e}\right)^2 - \frac{8}{3}\right]\frac{\omega^2T_{\nu}^2}{M_\mathrm{W}^4}(n_{\nu} - n_{\bar{\nu}}).
\end{equation}
Substituting into eq.~(\ref{Repko}) the relic neutrino asymmetry $| n_{\nu} - n_{\bar{\nu}}|=0.01T_{\nu}^3/6$, given by the dimensionless neutrino chemical potential $|\xi_{\nu}|=|\mu_{\nu}|/T=0.01$ at the present relic neutrino temperature $T_{\nu}\sim 2\,\mathrm{K}$, one gets the rotation angle,
\begin{equation}\label{Repko2}
|\phi|=8\times 10^{-56}\left(\frac{\omega}{{\rm eV}}\right)^2\left(\frac{l}{l_\mathrm{H}}\right)\,\mathrm{rad}.
\end{equation}
Applying eq.~(\ref{Repko2}) for extremely high energy photons with $\omega=10^{20}\,\mathrm{eV}$, in refs.~\cite{Abbasabadi:2001ps,Abbasabadi:2003uc}, the estimate $| \phi| \sim 8\times 10^{-16}\,\mathrm{rad}$ was obtained at the horizon size $l=l_\mathrm{H}=H_0^{-1}=4.3\times 10^3\,\mathrm{Mpc}$. For radiowaves, or even for optical photons $\omega\sim O({\rm eV})$, such a rotary power would be scanty. In optics, it is 40 orders of magnitude less, $|\phi|\sim 8\times 10^{-56}\,\mathrm{rad}$. 

Another result for extragalactic sources can be obtained for the neutrino-antineutrino sea embedded into isotropic plasma where transversal photons get the effective mass $\omega^2- k^2 =\omega_p^2\neq 0$ and we use our result in  eq.~(\ref{nonrel}),
\begin{equation}\label{nonrel2}
\frac{\phi}{l}=\Pi_2^{(\mathrm{T})}=- \frac{7\pi\alpha_\mathrm{em}}{3}(f_\mathrm{L}^0 - f_\mathrm{R}^0)\frac{n_e}{m_e^3}.
\end{equation}
For comparison we remind the result $\Pi_2^{(\mathrm{vac})}$ in eq.~(\ref{Mohanty2}) calculated for weak $\nu_e e$ interactions using one-loop Feynman diagrams with virtual electrons and $W$-bosons as in ref.~\cite{Mohanty:1997mr},
\begin{equation}\label{Mohanty4}
\frac{\phi}{l}=\Pi_2^{(\mathrm{vac})}= \frac{\sqrt{2}G_\mathrm{F}\alpha_\mathrm{em}}{3\pi}\left(\frac{\omega_p^2}{m_e^2}\right)(n_{\nu_e} - n_{\bar{\nu}_e}),
\end{equation}
where we formally substitute $q^2=\omega_p^2$.

First, both results do not depend on a photon frequency at all. For the plasma frequency $\omega_p=\sqrt{e^2n_e/m_e}=5.65\times 10^4\sqrt{n_e}\,\text{s}^{-1}=6.5\times 10^{-13}\,\mathrm{eV}$ given by a small $n_e\sim 3\times 10^{-4}\,\text{cm}^{-3}$ in the galo of a Milky-Way-like galaxy~\cite{Hammond:2012pn}, one obtains, at the distance $l=l_\mathrm{H}$, a very small value of the RM $|\phi|\simeq 8\times 10^{-44}\,\mathrm{rad}$. For the intergalactic electron density $n_e\sim 10^{-5}\,\text{cm}^{-3}$, this rotation angle would be lowered by a factor of 30.

In ref.~\cite{Mohanty:1997mr}, the rotary power in eq.~(\ref{Mohanty4}) was overestimated getting $\phi\sim 10^{-36}$, when the value $n_e=3\times 10^{-2}\,\text{cm}^{-3}$ was substituted. Moreover an archaic demand, that the neutrino energy density should not exceed the closure density of the universe, was used in ref.~\cite{Mohanty:1997mr}. It led to a great neutrino asymmetry density significantly exceeding the value $| n_{\nu}-n_{\bar{\nu}}|=0.01T^3_{\nu}/6$ for the neutrino asymmetry $|\xi_{\nu}|\simeq 0.01$.

Now let us turn to the analysis of the main result in eq.~(\ref{nonrel2}) rewritten as
\begin{equation}\label{our}
\frac{|\phi|}{l}=|\Pi_2^{(\mathrm{T})}| =\frac{7\pi\alpha_\mathrm{em}G_\mathrm{F}\sqrt{2}}{9}\left(\frac{n_e}{m_e^3}\right)0.244T_{\gamma}^3|\eta_{\nu}|,
\end{equation}
where we substituted for the factor $(f_\mathrm{L}^0 - f_\mathrm{R}^0)$ in eq.~(\ref{difference}) the total relic neutrino-antineutrino asymmetry density $\sum_a\Delta n_{\nu_a}=n_{\gamma}\eta_{\nu} =(0.244T_{\gamma}^3)\eta_{\nu} $ given by the relic photon density $n_{\gamma}=0.244 T_{\gamma}^3$. Equation~\eqref{our} gives $| \phi|=10^{-40}(l/l_\mathrm{H})\,\mathrm{rad}$ for the same electron density $n_e=3\times 10^{-4}\,\text{cm}^{-3}$. This value is three orders of magnitude greater than for $\nu\bar{\nu}$-gas in eq.~(\ref{Mohanty4}), that corresponds to the ratio of form factors $\Pi_2^{(\mathrm{T})}/\Pi_2^{(\mathrm{vac})}=7\pi/2\alpha_\mathrm{em}\sim 10^3$ in eqs.~(\ref{Mohanty3}) and~(\ref{nonrel}).

In eq.~(\ref{our}), we use the upper limit from WMAP+He bounds on the total neutrino asymmetry $\eta_{\nu}=\sum_a\eta_{\nu_a}$,  $a=e,\mu,\tau$, obtained in refs.~\cite{Castorina:2012md,Mangano:2011ip}: $-0.071<\eta_{\nu} < 0.054$. When flavors equilibrate owing to oscillations before BBN in the presence of a non-zero mixing angle (for the case $\sin^2\theta_{13}=0.04$ used in refs.~\cite{Castorina:2012md,Mangano:2011ip} involving all active neutrinos), the total neutrino asymmetry is distributed almost equally among the different flavors, leading to a final asymmetry, $\eta_{\nu_e}^{(\text{fin})}\approx\eta_{\nu_x}^{(\text{fin})}\approx \eta_{\nu}/3$, where $x=\mu, \tau$. We remind the definition of the asymmetry given by the partial degeneracy parameter $\xi_{\nu_a}\equiv \mu_{\nu_a}/T$,
\begin{equation}\label{eq:etadef}
  \eta_{\nu_a}\equiv \frac{n_{\nu_a} - n_{\bar{\nu}_a}}{n_{\gamma}}=\frac{1}{12\zeta(3)}\left[\pi^2\xi_a + \xi^3_a\right],
\end{equation}
which, accounting for $n_{\gamma}=2\zeta(3)T^3/\pi^2=0.244T^3$, gives for $\xi_{\nu_a}\ll 1$ the well-known fermion asymmetry density $\Delta n_{\nu_a}\equiv n_{\nu_a} - n_{\bar{\nu}_a}\simeq\xi_{\nu_a}T^3/6$. In eq.~\eqref{eq:etadef}, $\zeta(x)$ is the Riemann zeta function. Notice that the electron density in an intergalactic region $n_e$ is much less than the relic neutrino density at present, $n_e\ll \langle n_{\nu_a,\bar{\nu}_a}\rangle = 56\,\text{cm}^{-3}$. 

Note that, taking into account for the gravitational clustering of relic neutrinos in cold dark matter halos, one can expect an increase of the rotation angle given by eq.~(\ref{our}) due to a nonrelativistic neutrino (antineutrino) overdensity $n_{\nu, \bar{\nu}}/\langle n_{\nu, \bar{\nu}}\rangle\sim (10\div 100)$~\cite{Singh:2002de} obeying some upper bounds on the sum of the neutrino masses.

The Faraday rotation measure (RM) for the competing rotation of the polarization vector  at the angle $\phi=\lambda^2{\rm RM}$ in an intergalactic magnetic field (IGMF) $B_{\parallel}/\mu \mathrm{G}$, where $B_{\parallel}=({\bf B}\cdot {\bf k})/k$, is given by the well-known expression \cite{ZelRuzSok83} (see eq.~(\ref{index2}) below),
\begin{equation}\label{Faradayrotation}
{\rm RM}=0.81\times 4.3\times 10^{9}\left(\frac{n_e}{\text{cm}^{-3}}\right)\left(\frac{B_{\parallel}}{\mu \mathrm{G}}\right)\left(\frac{l}{l_\mathrm{H}}\right)\frac{{\rm rad}}{\text{m}^2},
\end{equation}
where we substitute $l_\mathrm{H}=4.3\times 10^9\,\text{pc}$~\cite{Rubakov}. Together with the estimate of the electron density in the intergalactic medium $n_e\simeq 3\times 10^{-4}\,\text{cm}^{-3}$ and the upper bound on IGMF $(B/\mu \mathrm{G}) < 10^{-3}$ known from CMB observations, eq.~\eqref{Faradayrotation} leads to the upper bound on RM,
\begin{equation}\label{upper}
|{\rm RM}|< 10^3\left(\frac{l}{l_\mathrm{H}}\right)~\frac{{\rm rad}}{\text{m}^2},
\end{equation} 
that corresponds to observable values ${\rm RM}=(\pm 10\div \pm 100)\,\mathrm{rad}/\text{m}^2$ for quasars in the radio wave band $\lambda\sim 1\,\text{m}$. 

\begin{figure}
  \centering
  \includegraphics[scale=0.5]{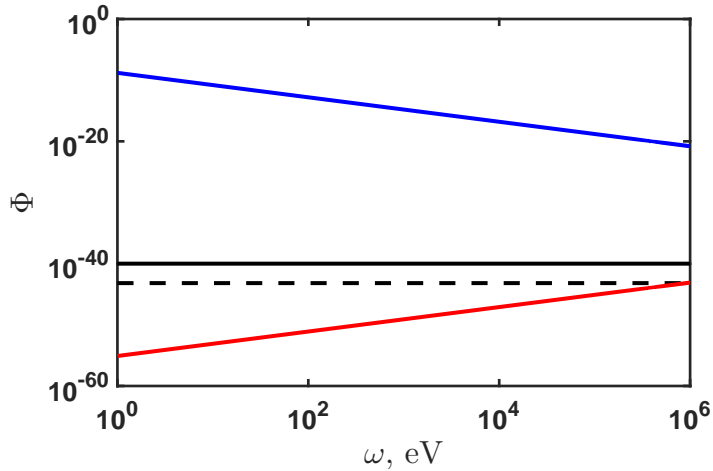}
  \caption{The rotation angle spectra $\Phi (\omega)=\phi (\omega)/(l/l_\mathrm{H})$ at the horizon size $l=l_\mathrm{H}$ for the birefringence of electromagnetic waves in the $\nu\bar{\nu}$ gas. The blue line corresponds to the hyperbolic dependence in eq.~(\ref{Faraday_bound}), $\phi\sim \omega^{-2}$, given by the Faraday rotation in IGMF with $B = 10^{-9}\,\mathrm{G}$ , $\phi=| \text{RM}|\lambda^2\sim \omega^{-2}$. The red line is the rotary power for the case of the $\nu\bar{\nu}$ gas in vacuum given by eq.~(\ref{Repko2}), $\phi\sim \omega^2$. The solid black line is our result in eq.~(\ref{our}) accounting for both an isotropic plasma and $\nu\bar{\nu}$ gas in intergalactic medium. The dashed black line is the vacuum contribution in eq.~(\ref{Mohanty4}) under the same conditions.\label{fig:rotarypower}}
\end{figure}

The bound in eq.~(\ref{upper}) gives the upper limit for the rotation angle in IGMF,
\begin{equation}\label{Faraday_bound}
|\phi|=|\mathrm{RM}| \lambda^2\leq 1.58\times 10^{-9}\left(\frac{l}{l_\mathrm{H}}\right)
  \left(
    \frac{\omega}{\text{eV}}
  \right)^{-2}\,\mathrm{rad}.
\end{equation}
Now, we can compare all estimates for rotation angles in different birefringence scenarios: $| \phi|\sim \omega^{-2}$ for the Faraday effect in IGMF, $|\phi|\sim \omega^2$ for the neutrino asymmetry in the $\nu\bar{\nu}$ gas in vacuum as given in eq.~(\ref{Repko2}) (with virtual charged leptons and $W$-bosons in Feynman diagrams), and our result in eq.~(\ref{our}), $|\phi|=10^{-40}(l/l_\mathrm{H})\,\mathrm{rad}=\text{const}$, that is independent of a frequency of EM waves. Such a comparison is illustrated
in figure~\ref{fig:rotarypower}.

\section{Discussion\label{sec:DISC}}

In the present work, we have reconsidered the possibility of the detection C$\nu$B using a birefringence of electromagnetic waves in the relic $\nu\bar{\nu}$ gas. In an intergalactic region at present, in addition to a sea of relic neutrinos and antineutrinos, there are a sparse isotropic plasma with the electron density $n_e\sim (3\times 10^{-4}\div 10^{-3})\,\text{cm}^{-3}\ll n_{\nu,\bar{\nu}}\sim 56\,\text{cm}^{-3}$,
and cosmological  IGMF's with the strength $B< 10^{-9}\,\mathrm{G}$. There are also relic photons and the Extragalactic Background Light (EBL) as a visible light from stars and photons in the infrared range from the  re-scattering of light on a dust in voids. In our study, we do not involve the two last ingredients with the densities $n_{\gamma}\simeq 400\,\text{cm}^{-3}$ and $n_\mathrm{EBL}\simeq 10^{-2}\div O(1)\,\text{cm}^{-3}\ll n_{\gamma}$. They are essential in the problem of a lower bound on IGMF~\cite{Neronov:1900zz}, when processes $\gamma + \gamma_\mathrm{EBL}\to e^+ + e^-$ and the inverse Compton scattering $e^{\pm} + \gamma_\mathrm{CMB}\to e^{\pm} + \gamma'$ are taken into account. In our scenario, the electromagnetic waves propagate from a remote source through the relic neutrino sea and a plasma, where IGMF is presented as well. We have estimated above how large different competitive birefringence effects are there; cf. figure~\ref{fig:rotarypower}.

Accounting for the surrounding plasma, for which off-shell photons with $q^2=\omega_p^2$ are applied, we consider additionally real charged particles instead of virtual charged leptons in Feynman diagrams, accounted for in refs.~\cite{Novikov,Abbasabadi:2001ps,Abbasabadi:2003uc,Mohanty:1997mr}, to calculate forward scattering amplitudes $\gamma\nu\to \gamma\nu$ and the corresponding refraction index for on-shell photons  $q^2=0$. In our case, one gets the new form factor $\Pi_2^{(\mathrm{T})}\sim G_\mathrm{F}$ in eq.~(\ref{Matsubara}), which is linear in the Fermi constant $G_\mathrm{F}$. This result is obtained in calculations of the relevant polarization operator in eq.~(\ref{interference}) using the Matsubara technique.

This form factor, namely its representation for the non-relativistic plasma in eq.~(\ref{nonrel}), is much greater than $\Pi_2^{(\nu)}$ in eq.~(\ref{Mohanty3}), $\Pi_2^{(\mathrm{T})}\gg \Pi_2^{(\nu)}$. Note that the term $\Pi_2^{(\nu)}$ in eq.~(\ref{Mohanty3}) stems from the correct substitution of the off-shell photon spectrum $q^2=\omega_p^2$ to the vacuum result derived in ref.~\cite{Mohanty:1997mr}. We also mention that the result of ref.~\cite{Mohanty:1997mr} coincides with eq.~(\ref{Mohanty2}) in the present work. Equation~(\ref{Mohanty2}) is obtained from eq.~(\ref{interference}) using the dimensional regularization method instead of the calculation of the Feynman diagram in figure~2(a) in ref.~\cite{Mohanty:1997mr}. Thus, the birefringence of EM waves in the relic $\nu\bar{\nu}$-gas embedded to plasma is given by the rotation of the polarization vector on the angle $\phi(l)=\Pi_2^{(\mathrm{T})}l$ in eq.~(\ref{angle}). It corresponds to the rotary power in eq.~(\ref{our}).

In the case of anisotropic plasma the refraction index for ordinary and extraordinary transversal EM waves in a  cold magnetized plasma reads
\begin{equation}\label{index}
n^{\pm}=\frac{k}{\omega}\sqrt{1 -\frac{\omega_p^2}{\omega (\omega \pm\omega_\mathrm{B})}},
\end{equation}
where $\omega_\mathrm{B}=eB_{\parallel}/m_e\simeq 6\times 10^{-17}\,\mathrm{eV}$ is the cyclotron frequency in the IGMF parallel to the photon momentum and $B_{\parallel}=({\bf B}\cdot{\bf k})/k\leq 10^{-9}\,\mathrm{G}$. Equation~\eqref{index} provides the birefringence in IGMF's [see eq.~(\ref{Faradayrotation})],
\begin{equation}\label{index2}
\phi= \omega(n^+ - n^-)l=\frac{\omega_p^2\omega_\mathrm{B}l}{2\omega^2}={\rm RM}\cdot\lambda^2.
\end{equation}
In eq.~(\ref{index2}), the cyclotron frequency $\omega_\mathrm{B}$ is much less than the plasma frequency $\omega_p=6.5\times 10^{-13}\,\mathrm{eV}$ in the intergalactic electron gas with the density $n_e\simeq 3\times 10^{-4}\,\text{cm}^{-3}$, $\omega_\mathrm{B}\ll \omega_p$.
The plasma frequency, in its turn, is much less, for instance, than a radio wave frequency in the GHz range, $\omega\simeq k= 4\times 10^{-6}\,\mathrm{eV}$, therefore $\omega_\mathrm{B}\ll \omega_p\ll \omega$. 

In the case of an isotropic plasma, $\omega_\mathrm{B}=0$, while $\omega_p^2\neq 0$, the birefringence effect is possible accounting for the electroweak $\nu e$ interaction in SM in the lowest order in $G_\mathrm{F}$, $\phi\sim G_\mathrm{F}$. We have obtained that the rotary power in eq.~(\ref{our}) is much larger than competing outcomes in refs.~\cite{Novikov,Abbasabadi:2001ps,Abbasabadi:2003uc} for EM frequencies below $\omega< O({\rm MeV})$; cf. figure~\ref{fig:rotarypower}.

Nevertheless, the predicted value of the rotary power (\ref{our}) remains negligible for observations by the present instruments at the background of $\Phi=\phi/l$ for EM waves in IGMF, which can be seen by comparing the black solid line and the blue line in figure~\ref{fig:rotarypower}. 

We do not extrapolate the curves $\Phi (\omega)$ in figure~\ref{fig:rotarypower} to more hard photon energies $\omega\gg {\rm MeV}$. Note that our outcome in eq.~(\ref{our}) becomes greater than the competing Faraday effect for IGMF $B_{\rm {IGMF}}/\mu{\rm G}\leq 10^{-3}$ in eq.~(\ref{Faraday_bound})  at huge photon energies $\omega\geq 4~{\rm PeV}$ remaining at the immeasurable level $|\phi|\sim 10^{-40}\,{\rm rad}$.

\appendix

\section{Dirac equation for electrons in the relic $\nu\bar{\nu}$ gas and polarization operator\label{Dirac}}

In the rest frame of $\nu\bar{\nu}$ gas, ${\bf f}_\mathrm{L,R}={\bf u}=0$, the Dirac equation for electrons $\mathrm{i}\partial_t\psi=H\psi$ is given by the Hamiltonian (see appendix~A in ref.~\cite{Dvornikov:2013bca}), 
\begin{equation}\label{Hamiltonian}
H=\bm{\alpha}{\bf p} + m\gamma_0 + V + V_\mathrm{A}\gamma^5,
\end{equation}
where $V$ and $V_\mathrm{A}$,
\begin{eqnarray}\label{potentials}
&&V= \frac{f_\mathrm{L}^0 + f_\mathrm{R}^0}{2}=G_\mathrm{F}\sqrt{2}\left[\sum_{\alpha}c_\mathrm{V}^{(\alpha)}\Delta n_{\nu_{\alpha}}\right],
\nonumber\\&&V_\mathrm{A}=-\frac{f_\mathrm{L}^0 - f_\mathrm{R}^0}{2}= G_\mathrm{F}\sqrt{2}\left[\sum_{\alpha}c_\mathrm{A}^{(\alpha)}\Delta n_{\nu_{\alpha}}\right],
\end{eqnarray}
are the vector and the axial potentials of the electroweak interaction. Here $c_\mathrm{V}^{(\alpha)}=2\xi \pm 0.5$, $c_\mathrm{A}^{(\alpha)}= \mp 0.5$ are the corresponding vector and axial constants for the ($\nu_{\alpha} e$) scattering in SM, where upper (lower) sign corresponds to $\nu_e e$  or $\nu_{\mu} e$ scatterings, and $\xi=\sin^2\theta_\mathrm{W}\simeq 0.23$ is the Weinberg parameter. 

In the rest frame for the relic neutrino gas, where $\mathbf {f}_\mathrm{L,R}=0$, one has for the number densities of massless neutrinos and antineutrinos
\begin{equation}\label{density}
n_{\nu_{\alpha},\bar{\nu}_{\alpha}}=\int \frac{\mathrm{d}^3p}{(2\pi)^3}\left[\exp\left(\frac{|{\bf p}| \mp \mu_{\nu_{\alpha}}}{T_{\nu_{\alpha}}}\right) + 1\right]^{-1},
\end{equation}
where $T_{\nu_{\alpha}}$ and $\mu_{\nu_{\alpha}}$ are the temperature and chemical potential for the $\alpha$ component of the $\nu\bar{\nu}$ gas, and $\alpha=e,\mu,\tau$.
In eq.~(\ref{potentials}), $\Delta n_{\nu_{\alpha}}=n_{\nu_{\alpha}}- n_{\bar{\nu}_{\alpha}}$ are the neutrino number density asymmetries.

In the rest frame of the QED plasma and the $\nu\bar{\nu}$ gas as a whole, $u^{\beta}=(1, {\bf 0})$, and
\begin{equation}\label{polarization}
\Pi_{\mu\nu}(q)=\Pi_{\mu\nu}^{(\mathrm{QED})}(q) + \mathrm{i}\varepsilon_{\mu\nu\alpha 0}q^{\alpha}\Pi_2(q),
\end{equation}
is the total polarization operator in SM.\footnote{Though the QED contribution is not important in the present task, we give its explicit form in the chosen frame of reference,
$$
\Pi_{\mu\nu}^{(\mathrm{QED})}(\omega, {\bf k})=q^2(\varepsilon_\mathrm{l}(\omega,k) - 1)e_{\mu}e_{\nu} + \omega^2(\varepsilon_\mathrm{tr}(\omega,k) - 1)\delta_{\mu i}\delta_{\nu j}(\delta_{ij} - \hat{k}_i\hat{k}_j),
$$
where $\varepsilon_\mathrm{l, tr}(\omega,k)$ are the longitudinal and transverse permittivities in an isotropic plasma that include small vacuum parts for the renormalized QED, $\omega^2\delta\varepsilon_\mathrm{tr}=q^2\delta\varepsilon_\mathrm{l}$ and $\delta\varepsilon_\mathrm{l,tr}\ll \varepsilon_\mathrm{l,tr}$. Here $e_{\mu}=(k,\omega \hat{k})/\sqrt{q^2}$ is the unit polarization vector $e_{\mu}e^{\mu}=-1$, obeying the condition $q^{\mu}e_{\mu}=0$, or $q^{\mu}\Pi^{(\mathrm{QED})}_{\mu\nu}=0$, as it should be.}
Here $q^{\alpha}=(\omega, {\bf k})$ is the photon momentum in medium, and the formfactor $\Pi_2(q)=(f_\mathrm{R}^0 - f_\mathrm{L}^0)\Pi_\mathrm{P}(q)$ provides the birefringence effect.  
The last parity-odd term in eq.~(\ref{polarization}), $\mathrm{i}\varepsilon_{\mu\nu\alpha\beta}q^{\alpha}u^{\beta}\Pi_2(q)$, stems from  the polarization operator accounting for electroweak interactions in the lowest Fermi approximation $\sim G_\mathrm{F}$,
\begin{equation}\label{interference}
\Pi_{\mu\nu}^{(\nu)}=\mathrm{i}e^2\int \frac{{\rm d}^4p}{(2\pi)^4}{\rm tr}\{\gamma_{\mu}S_0(p +q)\gamma_{\nu}S_1(p) + \gamma_{\nu}S_0(p)\gamma_{\mu}S_1(p +q)\}.
\end{equation}
Here $S_0(p)=(\gamma^{\mu}p_{\mu} + m)/(p^2 - m^2)$ is the standard charged lepton propagator and 
\begin{equation}\label{correction}
S_1(p)=-\frac{2V_\mathrm{A}}{p^2 - m^2}\left[\frac{\mathrm{i}\sigma_{\alpha\beta}\gamma^5p^{\alpha}u^{\beta}(\gamma^{\mu}p_{\mu} + m)}{p^2 - m^2}+ \frac{1}{2}\gamma_{\mu}\gamma^5u^{\mu}\right],
\end{equation}
is the term of the first order given by the axial potential $V_\mathrm{A}\sim G_\mathrm{F}$ in eq.~(\ref{potentials}) in the decomposition of the total propagator $S=S_0 + S_1 +\dotsb$ for a charged lepton interacting with a neutrino gas (see appendix~A in ref.~\cite{Dvornikov:2013bca}). 

The calculation of the parity-odd polarization operator in eq.~(\ref{interference}) is an alternative to the approach in refs.~\cite{Novikov,Abbasabadi:2001ps,Abbasabadi:2003uc} with the forward scattering amplitude for the reaction $\nu\gamma\to \nu\gamma$. Using the dimensional regularization method, one can calculate from the operator in eq.~(\ref{interference}) its vacuum part $\Pi_{\mu\nu}^{(\nu)}=-2\mathrm{i}V_\mathrm{A}\varepsilon_{\mu\nu\alpha\beta}q^{\alpha}u^{\beta}\Pi_\mathrm{P}^{(\mathrm{vac})}(q^2)$ with the form factor $\Pi_\mathrm{P}^{(\mathrm{vac})}(q^2)$ in eq.~(\ref{Mohanty}). Such result coincides with the outcome in ref.~\cite{Mohanty:1997mr} for the $\nu\bar{\nu}$ gas embedded in vacuum with zero temperature and chemical potentials, $T=\mu=0$, that is correct for virtual charged leptons and $W$-bosons in one-loop Feynmann diagrams for the $\nu\gamma$-scattering. 

In plasma with $T\neq 0$ and $\mu\neq 0$, the calculations in eq.~(\ref{interference}) in the integral $\int {\rm d}^4p(...)$ are given in ref.~\cite{Dvornikov:2013bca} using the Matsubara technique. For this purpose, after the replacement $\mathrm{i}\int {\rm d}p_0/2\pi\to T\sum_n$, $p_0=\mathrm{i}(2n + 1)\pi T + \mu$, $n=0,\pm 1, \pm 2,\dotsc$, the summation over the Matsubara frequencies is replaced by the integration in the complex plane with the inclusion of the Fermi distributions for real charged leptons. The computational details can be found, e.g., in  ref.~\cite{Kapusta}, as well as in appendix~C in ref.~\cite{Dvornikov:2013bca}.

For photons in empty space (without plasma) filled by a $\nu\bar{\nu}$ gas, the Chern-Simons Lagrangian term $\mathcal{L}_\mathrm{CS}=\Pi_2^{(\nu)}{\bf A}{\bf B}$ vanishes,  $\Pi_2^{(\nu)}=-2V_\mathrm{A}\Pi_\mathrm{P}^{(\mathrm{vac})}=0$ (see eq.~(\ref{Mohanty})). It provides the validity of the Gell-Mann theorem~\cite{Gell-Mann} for photons on shell, $q^2=\omega^2 - {\bf k}^2=0$.  However, in a plasma, the dispersion relations obey the condition $q^2\neq 0$. Thus, the birefringence effect is possible. Then, the plasma contribution corresponding to $T\neq 0$ and $\mu\neq 0$, prevails over the vacuum part $\Pi_2^{(\nu)}$ calculated assuming $T=\mu=0$, $\Pi_2^{(\mathrm{T})}\gg \Pi_2^{(\nu)}$, where $\Pi_2^{(\nu)}\sim q^2\neq 0$ [see eqs.~(\ref{Mohanty3}) and~(\ref{nonrel})].

\end{document}